\documentclass[twocolumn]{aastex631}
\usepackage{amsmath}
\usepackage{amssymb}
\usepackage{array}
\usepackage{float}
\usepackage{tabularx}
\usepackage{xcolor}
\usepackage{ulem}

\begin{document}

    \title{IC 10 X-1: A Double Black Hole Progenitor Probably Formed through Stable Mass Transfer}

    \author[0009-0002-3654-8775]{Gui-Yu Wang}
    \affiliation{Department of Astronomy, Nanjing University, Nanjing 210023, People's Republic of China}
    \affiliation{Key Laboratory of Modern Astronomy and Astrophysics, Nanjing University, Ministry of Education, Nanjing 210023, People's Republic of China}

    \author[0000-0003-2506-6906]{Yong Shao}
    \email{shaoyong@nju.edu.cn}
    \affiliation{Department of Astronomy, Nanjing University, Nanjing 210023, People's Republic of China}
    \affiliation{Key Laboratory of Modern Astronomy and Astrophysics, Nanjing University, Ministry of Education, Nanjing 210023, People's Republic of China}

    \author[0000-0003-3862-0726]{Jian-Guo He}
    \affiliation{Department of Astronomy, Nanjing University, Nanjing 210023, People's Republic of China}
    \affiliation{Key Laboratory of Modern Astronomy and Astrophysics, Nanjing University, Ministry of Education, Nanjing 210023, People's Republic of China}

    \author[0000-0002-3614-1070]{Xiao-Jie Xu}
    \affiliation{Department of Astronomy, Nanjing University, Nanjing 210023, People's Republic of China}
    \affiliation{Key Laboratory of Modern Astronomy and Astrophysics, Nanjing University, Ministry of Education, Nanjing 210023, People's Republic of China}

    \author[0000-0002-0584-8145]{Xiang-Dong Li}
    \affiliation{Department of Astronomy, Nanjing University, Nanjing 210023, People's Republic of China}
    \affiliation{Key Laboratory of Modern Astronomy and Astrophysics, Nanjing University, Ministry of Education, Nanjing 210023, People's Republic of China}

    \begin{abstract}
        IC 10 X-1 is one of close X-ray binaries containing a Wolf-Rayet donor, which can provide an evolutionary link between high-mass X-ray binaries and gravitational wave sources. It is still unclear about the precise nature of the accreting compact object in IC 10 X-1, although it looks more like a black hole than a neutron star. In this work, we use a binary population synthesis method to simulate the formation of IC 10 X-1 like binaries by assuming different common-envelope ejection efficiencies.
        This work represents a big step forward over previous studies since we adopt new criteria of mass-transfer stability. These criteria allow the formation of IC 10 X-1 like systems without experiencing common envelope evolution. Based on our calculations, we propose that the compact object in IC 10 X-1 is a black hole with mass of $\sim 10-30M_\odot$ and the progenitor evolution of this binary probably just experienced stable mass transfer.

    \end{abstract}

    \section{Introduction} \label{sec: introduction}
        IC 10 (also known as UGC 192) is a dwarf irregular galaxy that is currently undergoing active star formation \citep{Gholami2023}. 
        This galaxy is located in the Galactic plane (l=$118.97^{\circ}$, b=$-3.34^{\circ}$) with a distance of $\sim 0.7 \,\rm Mpc$ \citep{Borissova2000, Bauer2004}, making it the nearest starburst galaxy to the Milky Way. As the most luminous X-ray source in the galaxy IC 10 \citep{Brandt1997}, IC 10 X-1 is identified as an X-ray binary (XRB) containing a Wolf-Rayet (WR) star. The orbital period of this binary is $\sim 35\,\rm hr$, and the mass of the WR star is estimated to be between $17-35\, M_\odot$ \citep{Clark2004, Prestwich2007, Silverman2008}. Until now, however, the precise nature of the accreting compact object in IC 10 X-1 is still under debate. 

        In an XRB, the mass of the compact object can be dynamically measured according to the mass function. 
        Using the radial velocity curve obtained from the He II $\lambda4686$ emission line, \citet{Prestwich2007} suggested that the compact object in IC 10 X-1 is a massive black hole (BH) with mass of $23-34\, M_\odot$ \citep[see also][]{Silverman2008}. 
        Later, \citet{Laycock2015} argued that the mass estimate for IC 10 X-1 is invalid since the He II line is believed to originate from a shadowed region of the stellar wind rather than from the WR star itself. 
        It was further proposed that the compact object may be a low-mass BH or even a neutron star (NS). Based on \textit{Chandra} and \textit{NuSTAR} data, \citet{Steiner2016} demonstrated that the compact object in IC 10 X-1 is very likely to be a BH since the X-ray spectrum originated from the photosphere of an NS requires its radius much larger than those of known accreting NSs.
        Using the method of X-ray continuum fitting, the spin of this BH was constrained to be $\gtrsim 0.7$ if its mass is comparable to that in other known wind-fed BH XRBs. 
        Recently, it was suggested that the origin of the He II $\lambda4686$ emission line in IC 10 X-1 may be linked to multiple sources including a shadow region of the WR star and a hotspot located in the accretion disk \citep{Bhattacharya2023b, Bhattacharya2023a}. Overall, the mass measurement of the compact object in IC 10 X-1 remains unknown.

        When the WR star in IC 10 X-1 collapses into another compact object via a supernova explosion, the binary if survives is expected to become a gravitational wave (GW) source that will merge within the Hubble time \citep{Bulik2011}. After the discovery of the GW source GW150914 \citep{Abbott2016}, over 100 double compact object mergers have been identified and the majority of them are double BHs \citep{Abbott2023, Olsen2022, Nitz2023}. 
        The formation of these GW sources has been widely discussed involving various evolutionary channels \cite[see][for a review]{Mandel2022}. The popular channels include isolated binary evolution, chemically homogeneous evolution for rapidly rotating stars \citep[e.g.,][]{deMink2016,Marchant2016}, and dynamical interactions in star clusters \citep[e.g.,][]{Mapelli2022}, and etc. Given the evolution of isolated binary stars, two subchannels have been proposed according to the interactions between the compact object and its companion via either stable mass transfer \citep[SMT,][]{vandenHeuvel2017, Bavera2021a, Olejak2021, Shao2021, GG2023} or dynamically unstable mass transfer that followed by a common envelope (CE) phase \citep[e.g.,][]{Tutukov1993, Belczynski2016, Mapelli2018, Zevin2020}. Investigating the formation of IC 10 X-1 can potentially shed light on the possible channels of forming GW sources. 

        Assuming the compact object in IC 10 X-1 is a massive BH, \cite{Wong2014} simulated the formation of this binary via CE evolution. Using the standard \cite{Webbink1984} prescription to deal with CE evolution, they found that IC 10 X-1 is able to form with the need of a larger CE ejection efficiency than unity. Since this standard CE prescription leads to unphysical results, they employed a different treatment with the enthalpy formalism \citep{Ivanova2011} and then obtained that the CE ejection efficiency required to explain this system is in a reasonable range of $\sim 0.6-1$. Obviously, these results with different CE ejection efficiencies depend on the treatments of CE evolution, which are subject to big uncertainties \citep{Ivanova2013}. It was recently demonstrated for Roche-lobe filling binaries with a BH accretor that the process of mass transfer is more stable than previously thought \citep{Pavlovskii2017, Marchant2021}. In addition, \cite{vanSon2022} suggested that the formation of the double BH mergers with component masses above $30M_{\odot}$ is dominated by the SMT channel while the CE channel tends to produce the systems with less massive BHs. \cite{Shao2021} evolved a large gird of initial parameters for BH binaries with a nondegenerate donor to simulate the stability of mass transfer and obtained easy-to-use criteria to identify the occurrence of either SMT or CE evolution. Furthermore, their simulation showed that the binaries with heavier (lighter) BHs are more likely to undergo SMT (CE evolution). By adopting these new criteria, in this paper we revisit the formation of IC 10 X-1.

        The structure of this paper is organized as follows. 
        In Section \ref{sec: method}, we introduce our binary population synthesis (BPS) method along with observational constraints on the star formation history of the galaxy IC 10.
        The BPS outcomes under different formation channels are presented in Section \ref{sec: result}.
        In Section \ref{sec: discussion}, we make some discussions focusing on the origin of high-spin BHs, the observational sample of BH+WR binaries, and the key uncertainties that enter our simulations.
        We conclude in Section \ref{sec: conclusion}.

    \section{Method} \label{sec: method}
        \subsection{Population synthesis simulations with \texttt{BSE}} \label{sec: SECode}
            In this work, we use the \texttt{BSE} code \citep{Hurley2002} to simulate the formation of the binaries similar to IC 10 X-1 and obtain the population properties of these binaries. This code has been significantly modified including the treatment of mass loss due to stellar winds, the stability of mass transfer between binary components, and the mechanism of supernova explosions that form NSs and BHs \citep{Shao2014, Shao2021}. In the following, we list some important modifications relevant to the formation of IC 10 X-1 like binaries. 

            All binary models are computed starting from primordial binaries with two zero-age main-sequence stars. During the evolution of primordial binaries, we assume that mass transfer from the primary star to the secondary star is non-conservative and the accretion efficiency of the secondary star among all transferred matter is dependent on its rotational velocity \citep[]{Shao2014}. In this case, the criteria of mass-transfer stability, determining whether the primordial binaries undergo SMT or CE evolution, have been obtained by \citet{Shao2014} and incorporated in the \texttt{BSE} code. 

            After the hydrogen envelopes being stripped due to mass transfer, the primary stars finally collapse into compact objects of either NSs or BHs. 
            We adopt the delayed supernova mechanism \citep{Fryer2012} to deal with the masses of compact objects according to carbon-oxygen core masses before collapse.
            Under this mechanism, we assume that the kick velocities of natal NSs follow a Maxwellian distribution with a dispersion of $\sigma=265 \mathrm{~km} \mathrm{~s}^{-1}$ \citep{Hobbs2005}. For the kick velocities of natal BHs, we use the NS kick velocities reduced by a factor of $(1-f_{\rm fb})$, where $f_{\rm fb}$ is the fraction of the fallback material during supernova explosions. If the binaries are not disrupted due to supernova kicks, the survived systems contain a compact object and a non-degenerate star (the original secondary).

            When mass transfer occurs again in the binaries with a compact object and an evolved donor, we use the criteria recently obtained by \citet{Shao2021} to deal with mass-transfer stability. It is found that the binaries always experience SMT if the mass ratio of the donor to the compact object is less than the minimal value (i.e., $q < q_{\rm min} =2$), or always experience CE evolution if the mass ratio is larger than the maximal value (i.e., $q > q_{\rm max} =2.1+0.8M_{\rm co}$, where $M_{\rm co}$ is the mass of the compact object). For the binaries with the mass ratio between $q_{\rm min}$ and $q_{\rm max}$, whether they undergo SMT or CE evolution depends on the specific properties of the donor \citep[see Section 3 of][]{Shao2021}. These criteria are important for understanding the formation of IC 10 X-1 like binaries as post-mass transfer systems.

            During SMT, we assume that the outflow escaped from the binary systems carries away the specific orbital angular momenta of the accreting stars. For CE evolution, we use the standard energy formalism proposed by \citet{Webbink1984} to deal with the orbital evolution of the binary systems. This formalism includes the binding energy parameter $\lambda$ and the CE ejection efficiency $\alpha_{\rm CE}$. In our calculations, we employ the $\lambda$ values provided by \citet{Xu2010}. Since $\alpha_{\rm CE}$ represents the efficiency with which the orbital energy is used to eject the envelope of the donor, a reasonable $\alpha_{\mathrm{CE}}$ should be less than unity. 
            To explain the formation of IC 10 X-1, \cite{Wong2014} suggested a high CE ejection efficiency of $\alpha_{\mathrm{CE}} \geq 3.5$, or a reasonable value of less than unity when introducing the enthalpy formalism \citep{Ivanova2011}. 
            As a consequence, we adopt three different efficiencies of CE ejection  (i.e., $\alpha_{\mathrm{CE}}=1.0$, 3.0, and 5.0) to test their influence on the population of the formed binaries similar to IC 10 X-1.

        \subsection{Primordial Binaries} \label{sec: PB}
            The target sources containing a compact object and a WR star are the evolutionary products of primordial binaries. For a specific target source $i$, we can estimate its number by multiplying the formation rate $R_i$ by the time span $\triangle t_i$ during the WR stage. So the total number for all target sources from our calculations can be given as 
            \begin{equation} \label{eq: N}
                N = \sum_{i}(R_i \triangle t_i),
            \end{equation}
            where
            \begin{equation}
            R_i=\left(\frac{f_{\mathrm{bin}}}{2}\right)\left(\frac{\mathrm{SFR}}{M_*}\right) W_{b}
            \end{equation}
            depends on the parameter distribution of primordial binaries \citep[see also][]{Hurley2002}. Here, $f_{\mathrm{bin}}$ represents the binary fraction. 
            We set $f_{\mathrm{bin}}=1.0$, implying that all stars are born in binaries.
            SFR denotes the star formation rate of the galaxy IC 10. 
            Given the starburst history of IC 10 and the stellar evolution of the WR stars therein, \cite{Magrini2009} estimated that the star formation is more likely to have occurred within the last 10 Myr \citep[see also][]{Massey2007}. 
            In our calculations, we adopt a constant SFR of 0.5$M_{\odot} \mathrm{yr}^{-1}$ over the past 10 Myr, although the exact SFR is somewhat uncertain \citep{Leroy2006, Gholami2023}. In some cases, we extend the evolutionary time to 100 Myr. $\mathrm{M_*} \sim$ 0.5$M_{\odot}$ is the average mass of all stars.
            $W_{\mathrm{b}}$ weights the contribution of the specific binary evolved from a primordial binary, which is given by 
            \begin{equation}
                W_{\mathrm{b}}=\Phi\left(\ln M_1\right) \varphi\left(\ln M_2\right) \Psi(\ln a) \delta \ln M_1 \delta \ln M_2 \delta \ln a.
            \end{equation}
            Here $M_1$ is the primary mass, ranging from 5 to 100$M_{\odot}$ and following the relation of
            \begin{equation}
                \Phi\left(\ln M_1\right)=M_1 \xi\left(M_1\right),
            \end{equation}
            with 
              $\xi\left(M_1\right)$ describing the initial mass function  \citep{Kroupa1993}, as
            \begin{equation}
                \xi\left(M_1\right)=\left\{\begin{array}{cl}
                0 & M_1 \leqslant 0.1 M_{\odot} \\
                a_1 M_1^{-1.3} & 0.1 M_{\odot}<M_1 \leqslant 0.5 M_{\odot} \\
                a_2 M_1^{-2.2} & 0.5 M_{\odot}<M_1 \leqslant 1.0 M_{\odot} \\
                a_2 M_1^{-2.7} & 1.0 M_{\odot}<M_1<\infty
                \end{array}\right.,
            \end{equation}
            where $a_1 = 0.29056$ and $a_2 = 0.15571$ are the normalized parameters.
            And, $M_2$ is the secondary mass, ranging from 0.5 to 100$M_{\odot}$ and obeying a flat distribution between 0 and $M_1$ \citep{Kobulnicky2007}, i.e.,
            \begin{equation}
                \varphi\left(\ln M_2\right)=\frac{M_2}{M_1}.
            \end{equation}
            The orbital separation $a$ is taken from 3 to 10000R\textsubscript{$\odot$}, assumed to follow a uniform distribution in the logarithm \citep{Abt1983}, 
            \begin{equation}
                \Psi(\ln a)=k=\text{constant}.
            \end{equation}
            The normalization of this distribution gives $k = 0.12328$.
            In addition, we assume that all primordial binaries have circular orbits and set the metallicity Z = 0.2$\mathrm{Z_\odot}$ ($\mathrm{Z_\odot} = 0.02$) for all stars \citep{Leroy2006}. 
            With $\chi$ representing $M_1$, $M_2$, and $a$, the size of each interval $\delta\ln\chi$ in logarithmic space is determined as 
            \begin{equation}
            \delta \ln \chi=\frac{1}{n_{\chi}-1}(\ln \chi_{\rm max }-\ln \chi_{\rm min }),
            \end{equation}
            where $n_{\chi}$ denotes the number of grid points for a specific parameter $\chi$ and is set to $n^{1/3}$ (where $n$ is the number of the primordial binaries we evolved). In our calculations, we have simulated $10^7$ primordial binaries for each model. 

            From the evolutionary outcomes of primordial binaries, we pick out the target systems characterized by a BH or NS paired with a WR star, as potential candidates of IC 10 X-1. We record relevant parameters of these binaries and calculate their total number using Equation \ref{eq: N}. Although WR stars have typical masses of $\gtrsim 8M_\odot$ \citep{Crowther2007}, in this work we extend all helium stars with masses of $\geq2M_\odot$ to be WR stars.

\begin{figure*}[htbp]
    \centering
    \includegraphics[width=\textwidth]{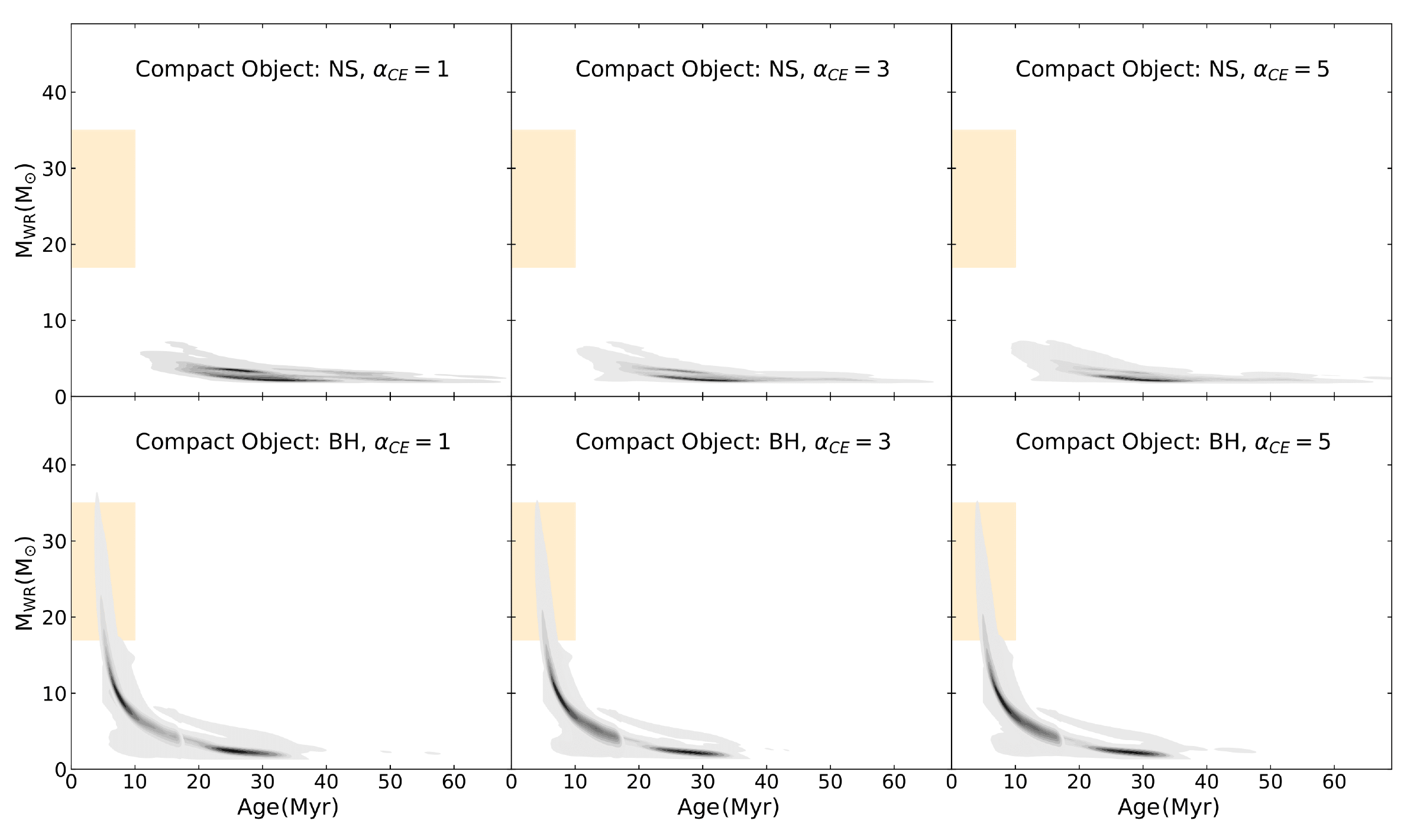}
    \caption{Predicted number distributions of the target binaries with a WR companion as a function of the WR mass ($M_{\rm WR}$) and the evolutionary age. In our calculations, the maximum evolutionary age is limited to 100 Myr. The top and bottom panels correspond to the compact object being an NS and a BH, respectively. The panels from left to right correspond to $\alpha_{\mathrm{CE}}$  increasing from 1 to 5.
    The orange rectangle in each panel denotes the region within which IC 10 X-1 is possible to form.}
    \label{fig: AgeMassNSBH}
\end{figure*}

    \section{Result} \label{sec: result}
        \subsection{The compact object: an NS or a BH?} \label{sec: CS}

            Observations of IC 10 X-1 indicated that the accretor is very likely to be a BH while it could also be an NS
        \citep{Laycock2015,Steiner2016}. 
            In this section, we aim to determine the nature of the accretor from an evolutionary point of view.
            Figure \ref{fig: AgeMassNSBH} presents the number distributions of the target binaries as a function of the WR mass and the evolutionary age. 
            The left, middle, and right panels denote the cases with $\alpha_{\mathrm{CE}}=1$, 3, and 5, respectively. 
            The top and bottom panels correspond to the compact object being an NS and a BH, respectively. 
            The orange rectangle covers the possible region of forming IC 10 X-1, which is plotted according to the IC 10 galaxy's star formation history of less than 10 Myr \citep{Massey2007} and the WR mass of 17$-$35$M_{\odot}$ \citep{Prestwich2007}. Overall, the total number of the target binaries containing an NS or a BH accretor is of the order of 10. 

            When the compact object is an NS, the progenitor evolution of the target binaries with a WR companion is expected to undergo a CE phase.
            In our models from  $\alpha_{\mathrm{CE}}=1$ to $\alpha_{\mathrm{CE}}=5$, the target binaries can only form within the age range of $\sim 10-70 $ Myr, and have the WR stars with masses of $\lesssim 9M_\odot$. These results are almost independent on the option of our adopted $\alpha_{\mathrm{CE}}$. On the one hand, formation of NSs requires relatively less-massive primary stars with stellar lifetimes longer than 10 Myr before core collapse. On the other hand, an NS is only able to successfully expel the envelope of a $\lesssim 20M_\odot$ donor star during CE evolution, so that the post-CE systems contain a $\lesssim 9M_\odot$ WR star. For the NS binaries with more massive donor stars, CE evolution always leads to merge even when taking $\alpha_{\mathrm{CE}}=5$. Obviously, we see that both the age and the WR mass for the NS+WR binaries from our calculations cannot match the observations of IC 10 X-1. 

            When the compact object is a BH, the progenitors of BH+WR binaries may experience SMT or CE phases during the evolution. Compared to NS+WR binaries, the target binaries with a BH accretor are relatively young systems with ages of $\sim 4-40 $ Myr and the masses of WR stars vary in the wide range of $\sim 2-37M_{\odot}$. 
            There is a tendency that the BH+WR binaries with longer ages have the WR stars with lower masses. Considering the constraints from the observations of IC 10 X-1, we estimate that the number of the BH+WR binaries within the region of the orange rectangle (see Figure \ref{fig: AgeMassNSBH}) is about $\sim 0.2-0.3$ when varying  $\alpha_{\mathrm{CE}}$ from 1 to 5. 

            Based on our calculations, we rule out the possibility that IC 10 X-1 contains an NS and conclude that it contains a BH. In the following analyses, we only consider the situation that IC 10 X-1 is a BH+WR binary. 

        \subsection{The Formation Channels} \label{sec: FC}

            Since IC 10 X-1 has a short orbital period of $\sim1.5$ days, some interactions are required to shrink the binary orbit during its progenitor evolution. The canonical formation channels of IC 10 X-1 like binaries (i.e., BH+WR binaries), as described in Section \ref{sec: SECode}, involve two mass-transfer interactions. The first mass transfer occurs from the primary star to the secondary star during the evolution of primordial binaries, and the second mass transfer happens from the original secondary to the BH (the descendant of the primary star). Depending on whether the mass transfer is stable or dynamically unstable, the main formation channels of BH+WR binaries are classified as follows.

            (1) The channel of SMT+SMT. Both mass transfer phases proceed stably.

            (2) The channel of SMT+CE. The first mass transfer phase proceeds stably, and the second mass transfer phase leads to CE evolution.

            (3) Other channels including CE+SMT, CE+CE and double-core CE. In these channels, the first mass transfer phase results in CE evolution. For the double-core CE channel, both components of the primordial binaries are required to have similar masses. As a consequence, both components can develop a compact core at the beginning of mass transfer and the post-CE binaries contain two naked helium cores \citep{Brown1995}.

\begin{figure}[htbp]
    \centering
    \begin{minipage}{0.4\textwidth}
    \includegraphics[width=\linewidth]{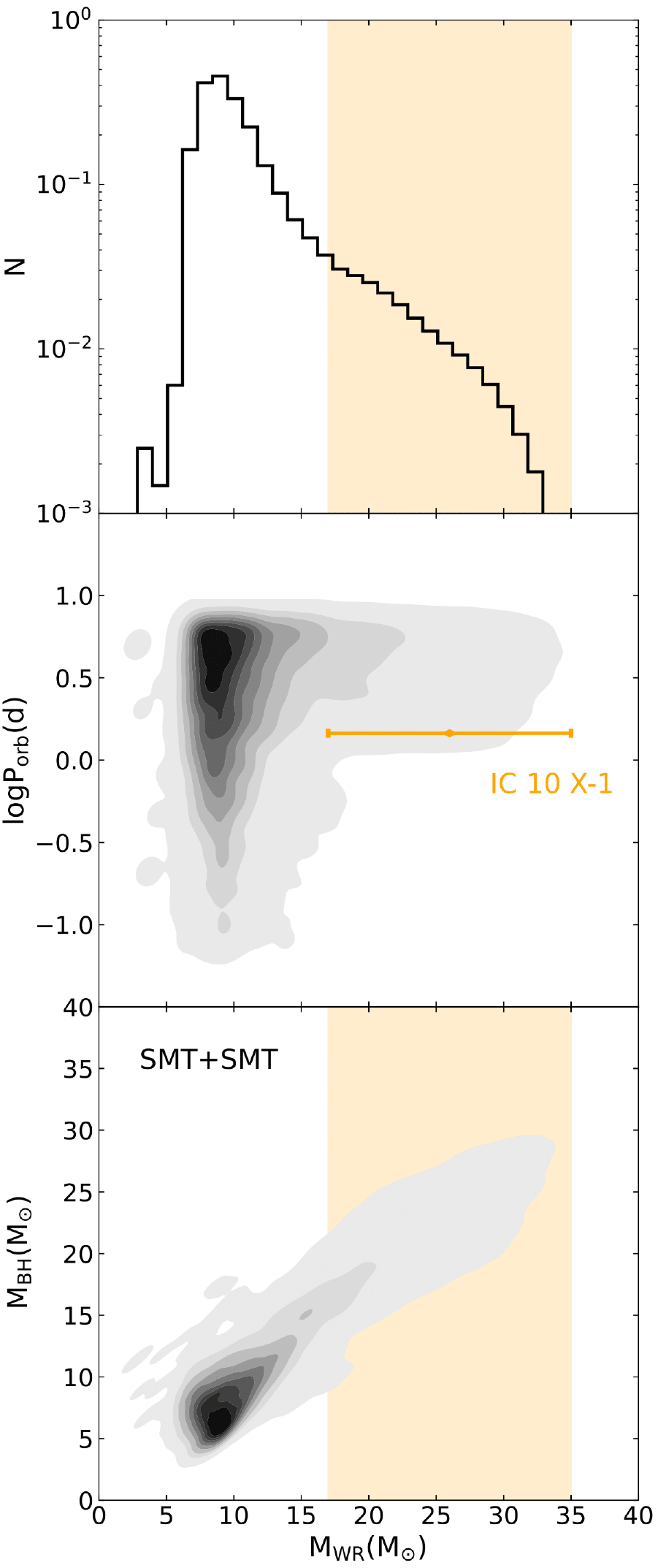}
    \end{minipage}%
    \caption{Predicted number distribution of the BH+WR binaries formed through the SMT+SMT channel, by assuming a constant star formation rate of 0.5 $M_{\odot} \mathrm{yr}^{-1}$ over the past 10 Myr for the galaxy IC 10. In the upper panel, the histogram is plotted showing the number distribution as a function of $M_{\mathrm{WR}}$. In the middle and lower panels, gray contours are plotted showing the number distribution in the $M_{\mathrm{WR}}-P_{\rm orb}$ and $M_{\mathrm{WR}}-M_{\mathrm{BH}}$ planes, respectively. 
    The orange dot marks the position of IC 10 X-1, and the orange rectangles correspond to its WR mass range of $17-35M_\odot$.}
    \label{fig: MassDALogPSingle}
\end{figure}

            \subsubsection{The channel of SMT+SMT} \label{sec: SMT}
                This channel allows the formation of the BH+WR binaries with long orbital periods up to $\sim 1000$ days \citep[see e.g.,][]{Shao2020}. 
                To clearly display the parameter distributions of the BH+WR binaries similar to IC 10 X-1, we only select the systems with orbital periods of less than 10 days. This threshold approximately corresponds to the maximum orbital period of $\sim 8.2$ days for M 101 X-1, among all BH+WR binaries known to date \citep{Liu2013, Esposito2015}.

\begin{figure}[htbp]
    \centering
    \begin{minipage}{0.4\textwidth}
    \includegraphics[width=\linewidth]{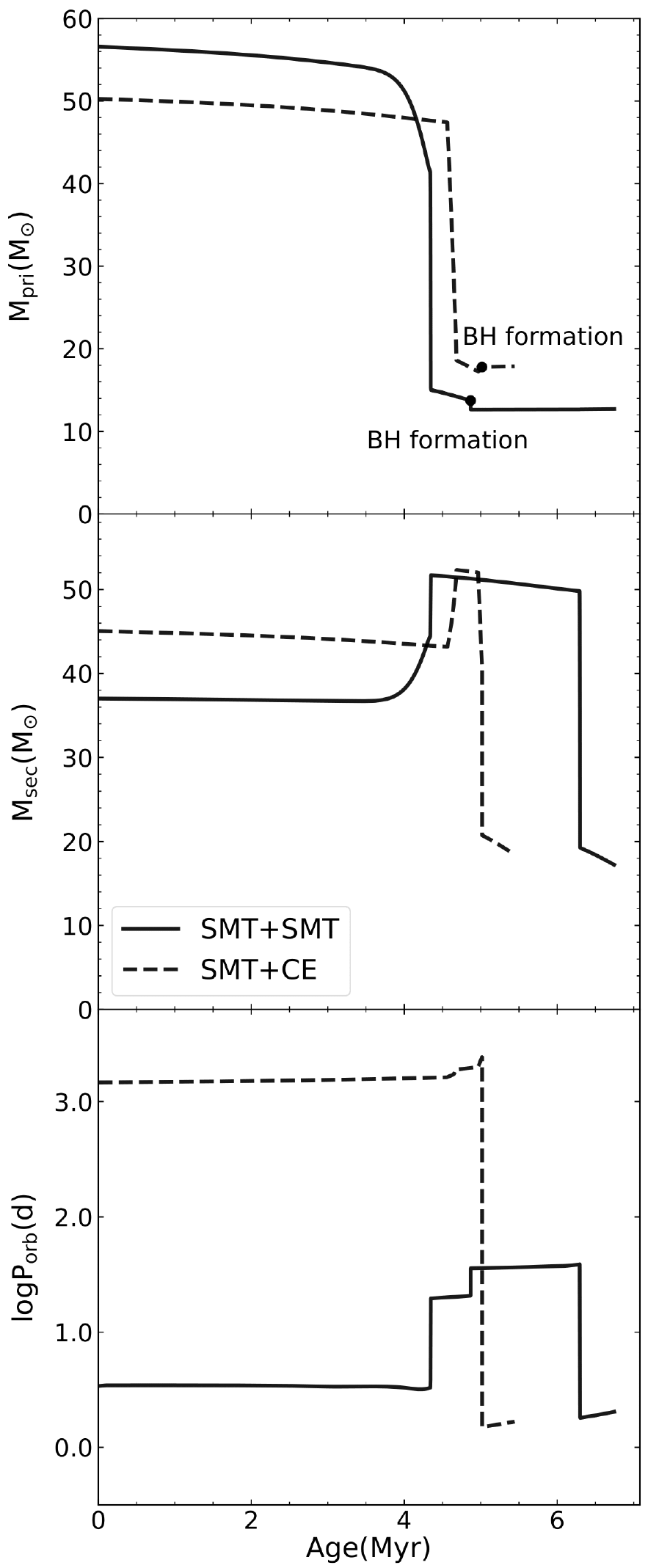}
    \end{minipage}%
    \caption{Evolution of two typical binaries that can lead to the formation of a system similar to IC 10 X-1. Panels from top to bottom correspond to the primary mass, the secondary mass, and the orbital period, respectively. The solid and dashed curves represent the SMT+SMT and SMT+CE channels, respectively. Two black dots mark the positions when a BH forms. During CE evolution, here we adopt $\alpha_{\mathrm{CE}}=5$.}
    \label{fig: m1process}
\end{figure}                 

\begin{figure*}[htbp]
    \centering
    \includegraphics[width=0.9\textwidth]{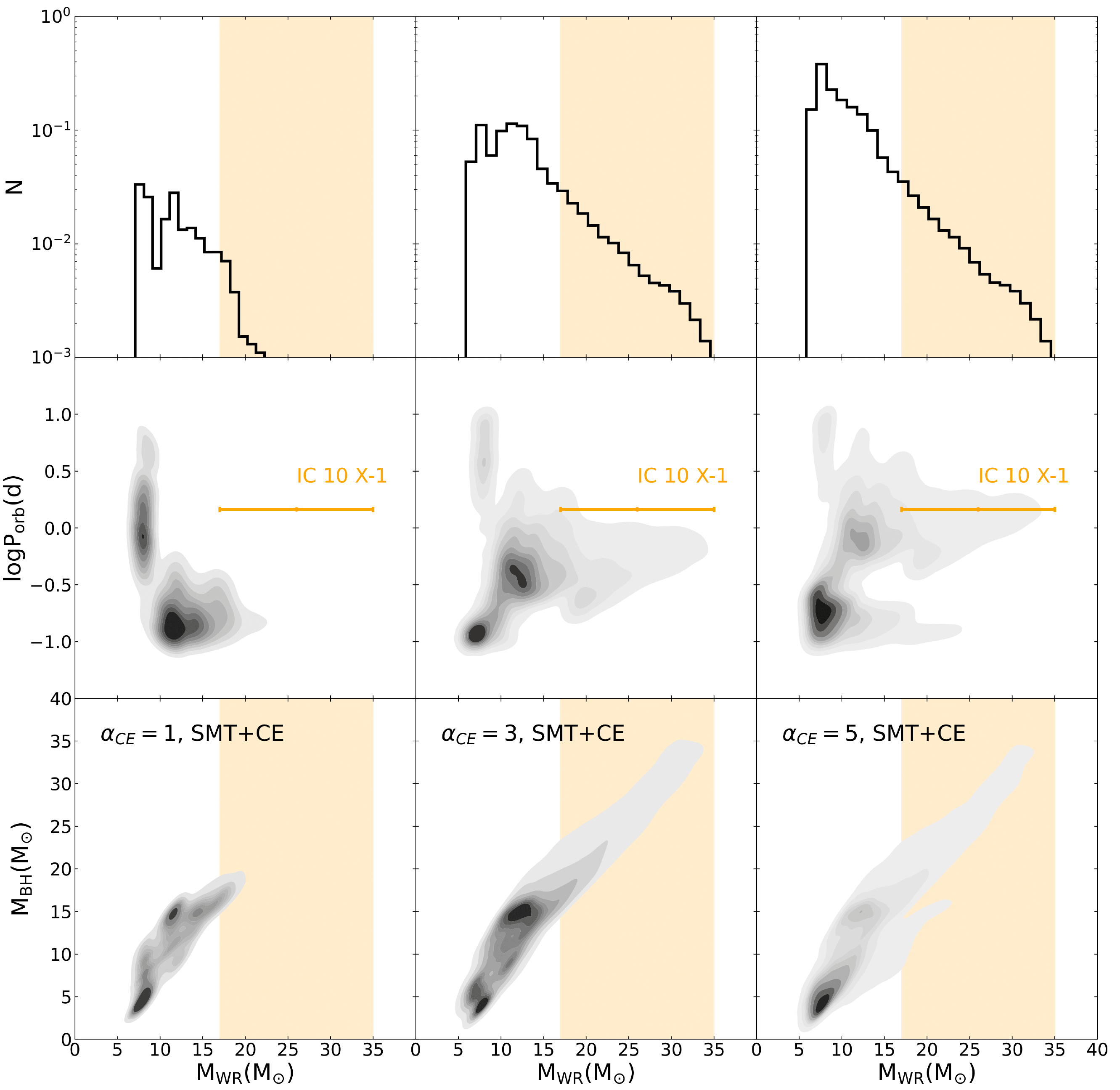}
    \caption{Similar to Figure \ref{fig: MassDALogPSingle}, but for the SMT+CE channel. The left, middle, and right panels correspond to the cases with 
    $\mathrm{\alpha_{\mathrm{CE}}}=$1, 3 and 5, respectively.}
    \label{fig: MassDALogPSMTCE}
\end{figure*}

                In the upper panel of Figure \ref{fig: MassDALogPSingle}, the histogram shows the number distribution of our obtained BH+WR binaries as a function of the WR mass. The orange rectangle marks the mass range of $17-35M_\odot$ for the WR star in IC 10 X-1. In the SMT+SMT channel, we expect that about 3 BH+WR binaries with orbital periods less than 10 days can form in the galaxy IC 10. Within the WR mass range of $ 17-35M_\odot$, the number of BH+WR binaries decreases to about 0.1.
                Moving to the middle and the lower panels, we depict the number distribution of IC 10 X-1 like binaries in the planes of $M_{\rm WR}-P_{\rm orb}$ and $M_{\rm WR}-M_{\rm BH}$, respectively. We can see that the BH+WR binaries formed via this channel cluster in the region with $M_{\rm WR}\sim 7-12M_\odot$ and $M_{\rm BH}\sim 3-10M_\odot$. There is a tendency that these binaries are more likely to contain low-mass components due to initial mass function. The orbital periods of these systems can reach as low as 0.1 days, although most of them have the orbital periods of $> 1$ day. Additionally, there exists nearly a positive correlation between $M_{\mathrm{BH}}$ and $M_{\mathrm{WR}}$. Although the binaries with $M_{\rm WR}= 17-35M_\odot$ are rare, our calculations can reproduce the observed features of IC 10 X-1 and predict that IC 10 X-1 should host a BH with mass of $\sim 10-30M_\odot$.        

                Figure \ref{fig: m1process} represents the evolutionary tracks (the solid curves) of a typical binary that can form an IC 10 X-1 like system via the SMT+SMT channel. The evolution starts from a primordial binary  with $M_{\mathrm{1}}\sim56M_{\odot}$ and $M_{\mathrm{2}}\sim37M_{\odot}$ in an orbit of $\sim$2.8 days. At the age of $\sim 4$ Myr, Case A mass transfer occurs from the primary to the secondary (both components are in the main sequence). After mass transfer, the primary becomes a WR star with mass of $\sim 15M_\odot$ and the secondary increases its mass to $\sim 52M_\odot$ due to accretion. At this moment, the orbital period of this binary is about 16 days. As a consequence, the averaged mass-transfer efficiency during this primordial binary evolution is $\sim 0.4$. At the age of $\sim 4.86$ Myr, the WR star directly collapses into a BH with $M_{\rm BH}\sim 12 M_\odot$ \citep[during which about $1M_\odot$ of baryonic mass is reduced via neutrinos that are lost,][]{Timmes1996}, and the orbital parameters of the binary change immediately. After $\sim 6.3$ Myr, the binary experienced the second SMT evolves to be a BH+WR system with $M_{\mathrm{BH}}\sim14M_{\odot}$ and $M_{\mathrm{WR}}\sim19M_{\odot}$ in a 1.5-day period. This stage lasts about 0.5 Myr until the second WR star collapses.

                In the SMT+SMT channel, IC 10 X-1 is the evolutionary product of BH XRBs with a supergiant companion that experienced an SMT phase. After reanalyzing our recorded data, we obtain that the progenitor binaries of IC 10 X-1 have a $\sim 36-74M_\odot$ supergiant with an orbital period of $\sim 2-30$ days. It is worth noting that some known BH XRBs with O-type supergiant companions resemble the configurations of IC 10 X-1's progenitor binaries. 
                Cygnus X-1 contains a $40.6^{+7.7}_{-7.1}M_{\odot}$ supergiant in a 5.6-day orbit \citep{Miller2021}, M33 X-7 hosts a $70.6\pm6.9 M_{\odot}$ supergiant in a 3.45-day orbit \citep{Orosz2007}, and LMC X-1 has a $31.79\pm3.48 M_{\odot}$ supergiant in a 3.9-day orbit \citep{Orosz2009}.
                The BHs in these observed XRBs are estimated to have the mass of $\sim 10-20M_\odot$, which is consistent with our predicted BH mass for IC 10 X-1. 

            \subsubsection{The channel of SMT+CE} \label{sec: CE}
                In this channel, the second mass-transfer phase leads to CE evolution. As a result, the majority of the BH+WR binaries from our calculations are close systems with orbital periods of $\sim 0.1-1$ days. In Figure \ref{fig: MassDALogPSMTCE}, we depict the number distribution of the BH+WR binaries formed via the channel of SMT+CE. These systems have the WR masses in the range of $\sim 6-35 M_{\odot}$ and the BH masses of $\sim 3-30 M_{\odot}$. 
                The left, middle, and right panels denote the cases with $\alpha_{\mathrm{CE}}=1$, 3, and 5, respectively. In the $\alpha_{\mathrm{CE}}=1$ case, the BH+WR binaries are mainly distributed in two regions with relatively high component masses and short orbital periods or relatively low component masses and long orbital periods, because the magnitudes of the binding energy parameter of CE evolution are sensitive to the masses and the evolutionary stages of donor stars \citep[e.g.,][]{Xu2010,Wang2016}. Varying $\alpha_{\mathrm{CE}}$ from 1 to 5 results in the formation of the BH+WR binaries with increasing orbital periods. Meanwhile, some systems that originally merged in the $\alpha_{\mathrm{CE}}=1$ case can survive CE evolution when $\alpha_{\mathrm{CE}}\geq 3$, leading to the emergence of additional substructures in the corresponding contour diagram.

                From the left panel of Figure \ref{fig: MassDALogPSMTCE}, corresponding to the case of $\alpha_{\mathrm{CE}}=1$, we see that our calculations cannot reproduce the BH+WR binaries with the same properties as IC 10 X-1. Overall, the number of our obtained BH+WR binaries is totally about 0.2. For the systems with $M_{\rm WR} = 17-35M_\odot$, their number decreases to 0.02 and they are expected to be have the orbital period of $\sim 0.1-0.3$ days which significantly smaller than that of IC 10 X-1. 

                For the middle panel representing the case of $\alpha_{\mathrm{CE}}=3$,
                we expect that there are totally about 0.9 BH+WR binaries. Among them, $\sim 0.1$  systems have a $ 17-35M_\odot$ WR donor. And, these systems have the orbital period of $\lesssim 1.5$ day which still lower than that of IC 10 X-1. So we demonstrate that IC 10 X-1 can still hardly form in this case. 

                For the right panel, representing the case of $\alpha_{\mathrm{CE}}=5$,
                we estimate that the number of all BH+WR binaries formed in the SMT+CE channel is $\sim 1.6$, and decreases to $\sim 0.2$ for the systems with a $17-35M_\odot$ WR donor. Only when adopting a high CE ejection efficiency of $\alpha_{\mathrm{CE}}=5$, our calculated results can well match the observations of IC 10 X-1 including its WR mass and orbital period. In this case, we find that IC 10 X-1 should also possess a $\sim 10-30M_\odot$ BH.

                In the SMT+CE channel, we select a typical binary from our calculations that can evolve to be an IC 10 X-1 like system. The dashed curves in Figure \ref{fig: m1process} denote the evolutionary tracks of this binary as a function of age. The primordial binary comprises a primary star with $M_{1}\sim51M_{\odot}$ and a secondary star with $M_{2}\sim45M_{\odot}$ in an orbit of $\sim1467$ days. After $\sim4.5$ Myr, the binary experiences an SMT via Case B Roche lobe overflow. The post-mass transfer system has a WR star of $\sim19M_{\odot}$ and an O-type star of $\sim52M_{\odot}$. We can estimate that the averaged mass-transfer efficiency is $\sim 0.2$ during this SMT phase. Then, the WR star collapses into a BH with $M_{\rm BH}\sim 18M_\odot$. At the age of $\sim 5$ Myr, the second mass transfer phase leads to CE evolution and the binary orbit greatly decreases from $\sim 2000$ days to $\sim 1.5$ days. Finally, the binary becomes a BH+WR system with  $M_{\mathrm{BH}}\sim18M_{\odot}$ and $M_{\mathrm{WR}}\sim21M_{\odot}$.

            \subsubsection{Other channels} \label{sec: other}
\begin{figure*}[htbp]
    \centering
    \includegraphics[width=0.9\textwidth]{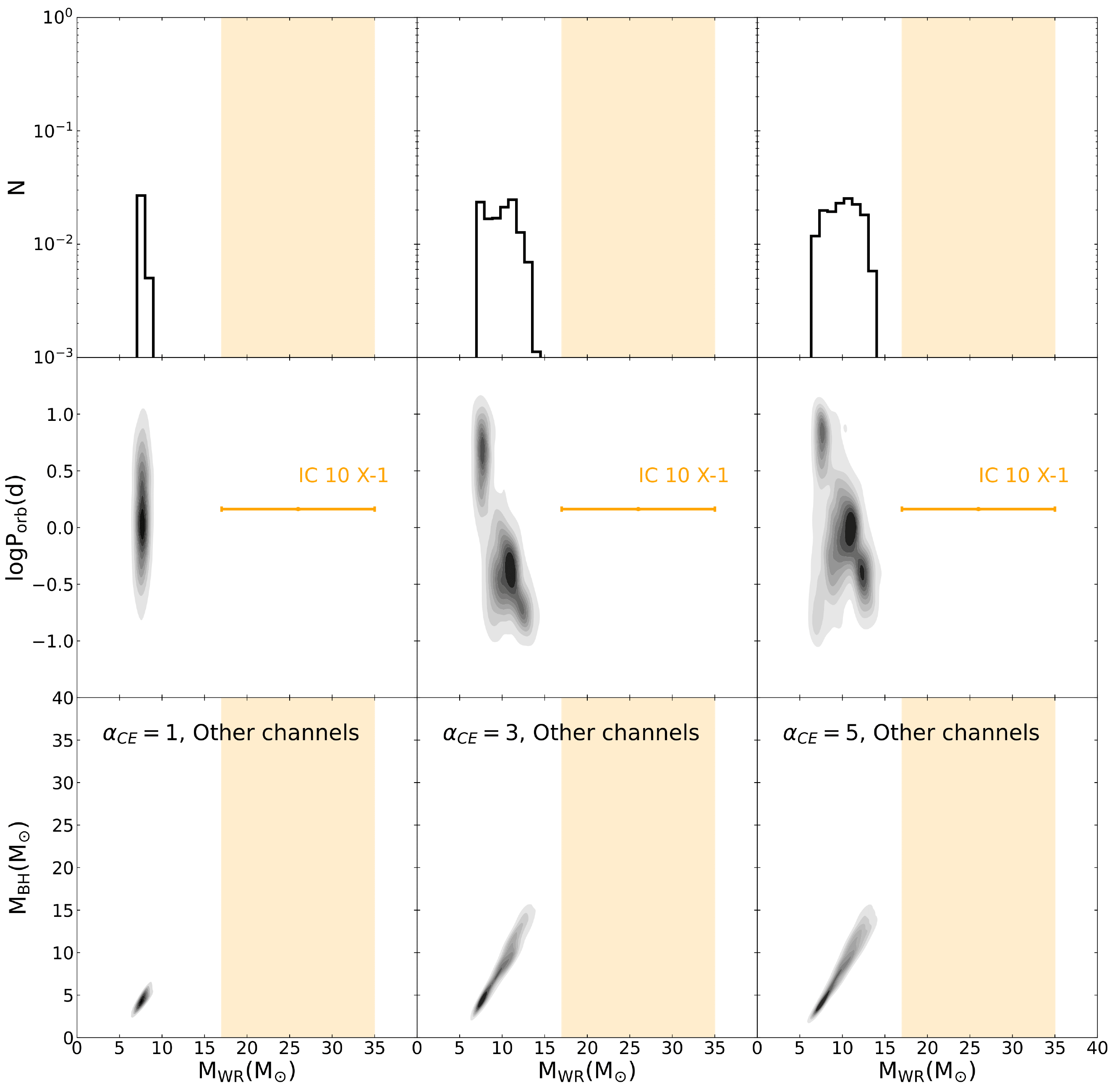}
    \caption{Similar to Figure \ref{fig: MassDALogPSMTCE}, but for other formation channels excluding SMT+SMT and SMT+CE.}
    \label{fig: MassDALogPOther}
\end{figure*}

                Excluding SMT+SMT and SMT+CE, we consider other channels to form BH+WR binaries and show their number distribution in Figure \ref{fig: MassDALogPOther}. As mentioned before, other channels have a common feature that the first mass-transfer phase results in CE evolution. When changing $\alpha_{\mathrm{CE}}$ from 1 to 5, our calculations show that only $\sim 0.03-0.1$ BH+WR binaries can from via other channels and they are expected to have the WR stars with masses of $\sim 6-15M_\odot$. 
                Therefore, other channels cannot explain the formation of IC 10 X-1. 

                We conclude that IC 10 X-1 probably originates from massive binary evolution without experiencing any CE phase, i.e., via the SMT+SMT channel. Alternatively, the SMT+CE channel is also able to produce IC 10 X-1 if assuming $\alpha_{\mathrm{CE}} \gtrsim 3$ \citep[see also][]{Wong2014}.

    \section{Discussion} \label{sec: discussion}
        \subsection{The high spin of IC 10 X-1's BH} \label{sec: hsBH}
            \cite{Steiner2016} suggested that IC 10 X-1 may contain a high-spin BH with $a_{*}>0.7$. IC 10 X-1 is a BH+WR binary that building an evolutionary link between XRBs with O-type supergiant companions and double BH mergers as GW sources. For the former binaries, the  BHs are likely to have high spins of $\gtrsim 0.8$ \citep[e.g.,][]{Qin2019, Draghis2023}. GW observations of double BH mergers indicate that BH's spin magnitudes appear concentrated below $a_* \lesssim 0.4$, with a possible tail extending toward large values \citep{Abbott2023}. One possibility for this discrepancy is that a small fraction of double BH mergers experienced a similar evolutionary pathway to the observed XRBs with a high-spin BH \citep{Fishbach2022}.
            Recently, \citet{Zdziarski2024} argued that BH spin measurements in BH+O XRBs are highly sensitive to the assumptions on the structure of the accretion disk around the BH, and pointed out that BH's spins could be relatively low and only weakly constrained. Their result provided an alternatively possible explanation of the tension between the BH spins inferred from the observations of GW sources and BH+O XRBs.

            The origin of BH spins involves the increase of angular momenta from the collapsing cores before BH formation, or from the mass accretion after BH formation \citep{Belczynski2020}. If the BH of IC 10 X-1 indeed has a high spin of $a_*>0.7$, we discuss the formation channel of this binary.
            In the SMT+CE channel,
            tidal interaction acted on the BH's progenitor is inefficient to produce the high spin of this BH since the post-SMT binaries have wide orbits with $P_{\rm orb}\gg 1$ day \citep{Belczynski2020, Bavera2021b}. During the subsequent CE evolution, the BH can hardly accrete to increase its spin. In this channel, IC 10 X-1's BH is more likely to have a low spin as observed in most of double BH mergers. In the SMT+SMT channel, assuming weak tidal coupling between the stellar core and the envelope is able to produce high-spin BHs if the first SMT phase occurs in close binaries with Case A mass transfer \citep[as the BHs observed in BH+O XRBs,][]{Qin2019}. Furthermore, stable mass accretion during the second SMT phase may significantly increase BH's spins \citep{Shao2022a,Briel2023,Xing2024}. As a consequence, the SMT+SMT channel has an advantage to explain the high spin of the BH in IC 10 X-1.

        \subsection{Observational sample of BH+WR binaries} \label{sec: othsr}
            Besides IC 10 X-1, there are some other BH+WR binaries observed in the Milky Way and nearby galaxies. Cyg X-3 is the only known XRB containing a WR donor in the Milky Way and this binary has a close orbit with $P_{\rm orb} \sim 0.2$ days \citep{vanKerkwijk1992}. The masses of the WR star and the compact object were estimated by \citet{Zdziarski2013} to be $10.3^{+3.9}_{-2.8}M_\odot$ and $2.4^{+2.1}_{-1.1}M_\odot$, respectively. The mass range of the compact object allows it to be either an NS or a BH. Analyses of observational data suggested that this compact object is very likely to be a BH \citep{Zdziarski2013, Antokhin2022}. 
            NGC 300 X-1 contains a BH with mass of $17^{+4}_{-4}M_\odot$ and a WR star with mass of $26^{+7}_{-5}M_\odot$ in a 1.4-day orbit \citep{Carpano2007, Carpano2019, Binder2021}, which has very similar properties to IC 10 X-1. 
            M 101 ULX-1 consists of a $\sim 5 - 30 M_\odot$ BH  and a $19^{+1}_{-1} M_\odot$ WR star in a  8.2-day orbit \citep{Liu2013}. In addition, there are several BH+WR candidates with  undetermined component masses \citep[e.g.,][]{Esposito2015,Lin2023}.

            In the Appendix \ref{appendix}, we present calculated number distributions of the BH+WR binaries formed from different channels. Here, we relax the maximum evolutionary age of BH+WR binaries to be 100 Myr. This allows the formation of more BH+WR binaries with a relatively low-mass ($\lesssim 8M_\odot$) WR star\footnote{In contrast to the observations of the BH+WR binaries with $M_{\rm WR}\gtrsim 8M_\odot$, our calculations show that most of systems have a low-mass WR donor. The detection of BH+WR binaries may be subject to some observational biases \citep[e.g.,][]{Sen2021,Sen2024}, which are not considered in this work.}. Based on our calculations, we see that only the SMT+SMT channel can explain the formation of M 101 ULX-1 since it has a long orbital period. For NGC 300 X-1, its orbital period and component masses can be reproduced via the SMT+SMT channel and the SMT+CE channel with $\alpha_{\rm CE}=5$. Previous works have demonstrated that Cyg X-3 is able to form in the SMT+CE channel with $\alpha_{\rm CE}=1$ \citep[e.g.,][]{Shao2020}, which is consistent with our results. Interestingly, we find that the SMT+SMT channel may also lead to the formation of Cyg X-3, as adopting the new criteria of mass-transfer stability in our current work. 

\begin{figure}[htbp]
    \centering
    \begin{minipage}{0.4\textwidth}
    \includegraphics[width=\linewidth]{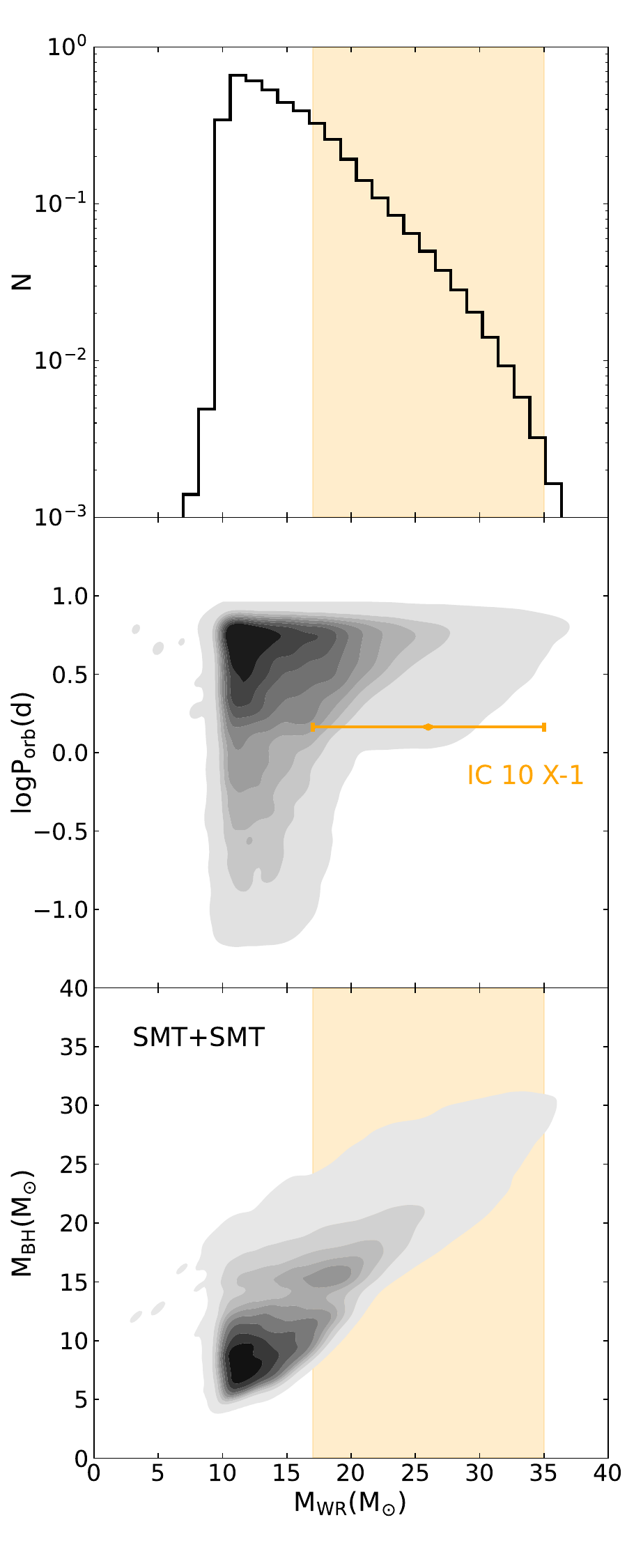}
    \end{minipage}%
    \caption{Similar to Figure \ref{fig: MassDALogPSingle}, but assuming the mass-transfer efficiency during primordial binary evolution to be 0.5.}
    \label{fig: SS_a5_m2}
\end{figure}

\begin{figure*}[htbp]
    \centering
    \includegraphics[width=0.9\textwidth]{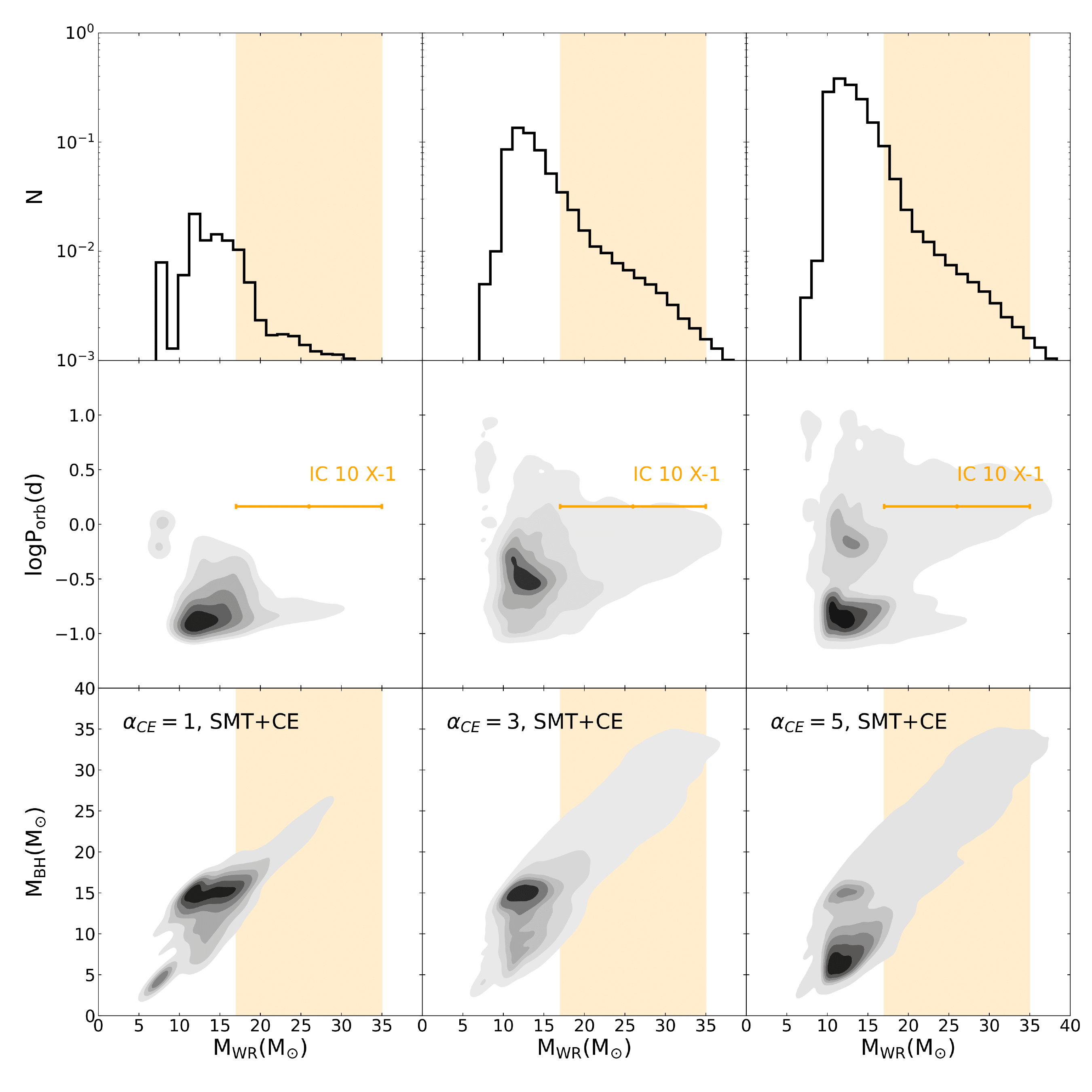}
    \caption{Similar to Figure \ref{fig: MassDALogPSMTCE}, but assuming the mass-transfer efficiency during primordial binary evolution to be 0.5.}
    \label{fig: SC_a5_m2}
\end{figure*}

        \subsection{Mass transfer efficiencies during primordial binary evolution} \label{sec: ame}
            In our calculations, we have assumed that the mass-transfer efficiencies that the accreted matter by the secondary among all transferred matter from the primary depend on the rotational velocities of the secondary. Under this assumption, the secondary can hardly accrete any material if it rotates close to its critical limit. This rotation-dependent picture of mass-transfer efficiency can well explain the formation of Galactic WR+O binaries \citep{Petrovic2005,Shao2016}. In addition, \citet{Wellstein2001} presented the evolutionary calculations of massive binaries with conservative mass transfer and found that none of observed WR+O binaries can fit their calculated results. On the other hand, the mass distribution of the Be stars with an NS companion suggests a mass-transfer efficiency of $\geq 0.3$ during the progenitor evolution \citep{Shao2014,Vinciguerra2020}. In this section, we assume that half of the transferred mass is accreted by the secondary and the other half is lost from the system \citep[i.e., Model II of][]{Shao2014}, and test the influence of mass-transfer efficiencies on the formation of IC 10 X-1.

            When assuming the mass-transfer efficiency to be 0.5, Figure \ref{fig: SS_a5_m2} represents the predicted number distribution of the BH+WR binaries formed through the SMT+SMT channel. It can be seen that these systems have more massive components with $M_{\rm WR}\gtrsim 10M_\odot$ and $M_{\rm BH}\gtrsim 5M_\odot$, compared to the rotation-dependent assumption of mass-transfer efficiency. And, the number of the BH+WR systems with $M_{\rm WR}=17-35M_\odot$ increases to 0.6. Similarly, the mass of the BH in IC 10 X-1 is expected to be in the range of $\sim 10-30M_\odot$.

            In Figure \ref{fig: SC_a5_m2}, we show the number distribution of the BH+WR binaries formed via the SMT+CE channel by assuming a mass-transfer efficiency of 0.5. As expected, this assumption significantly reduces the formation of the systems with $M_{\rm WR}\sim 6-10M_\odot$ and $M_{\rm BH}\sim 3-5M_\odot$. When changing $\alpha_{\mathrm{CE}}$ from 1 to 5, the number of the BH+WR binaries with $M_{\rm WR}=17-35M_\odot$ increases from 0.03 to 0.2. In the cases with  $\alpha_{\mathrm{CE}}=3$ and $\alpha_{\mathrm{CE}}=5$, our calculations can match the IC 10 X-1's observations of both the WR mass and the orbital period.
        
            Besides mass-transfer efficiencies, there exist other uncertain factors during binary evolution, which are also able to influence the relative contributions of IC 10 X-1 like systems formed via the SMT+SMT and SMT+CE channels. For example, the recently updated modelling of stellar winds \citep{Vink2024} and supernova explosions \citep{Schneider2021} is bound to affect the masses of remnant compact objects.


    \section{Conclusion} \label{sec: conclusion}
        We have employed a BPS method to investigate the formation channels of IC 10 X-1 and the nature of the compact object therein, by incorporating recently updated criteria to deal with mass-transfer stability in binaries with a BH accretor \citep{Shao2021}. It is assumed that the starburst galaxy IC 10 undergoes a constant SFR of $0.5M_\odot\,\rm yr^{-1}$ over the past 10 Myr. Varying CE ejection efficiencies from 1 to 5, we obtain that all NS+WR binaries in the galaxy IC 10 host a WR star with mass of $\lesssim9M_\odot$. This result is in conflict with the observations of IC 10 X-1. Our calculations support the suggestion of previous works that the compact object is a BH.
        Considering the observational constraints from the WR mass and the orbital period, we demonstrate that the SMT+SMT channel can naturally explain the formation of IC 10 X-1. In the case of $\alpha_{\rm CE}\gtrsim3$, IC 10 X-1 can also form via the SMT+CE channel. In both channels, we expect that the galaxy IC 10 possesses $\sim 0.1$ BH+WR binaries with properties similar to IC 10 X-1. Importantly, the SMT+SMT channel is more likely to produce a high-spin BH as observed in IC 10 X-1. In addition, the overall parameter distributions of the observational sample of BH+WR binaries also favor the SMT+SMT channel. Our work emphasizes that the SMT+SMT channel is important for the formation of close BH+WR binaries as the progenitors of double BH mergers and the understanding of the origin of high-spin BHs.

    \vspace{20pt}
    We thank the referee for constructive suggestions that helped improve this paper. This work was supported by the National Key Research and Development Program of China (Grant Nos 2023YFA1607902 and 2021YFA0718500), the Natural Science Foundation of China (Nos 12041301, 12121003, and 12373034), the Strategic Priority Research Program of the Chinese Academy of Sciences (Grant No. XDB0550300).

    \appendix

\section{Number distribution of BH+WR binaries}\label{appendix}

\begin{figure}
    \centering
    \begin{minipage}{0.4\textwidth}
    \includegraphics[width=\linewidth]{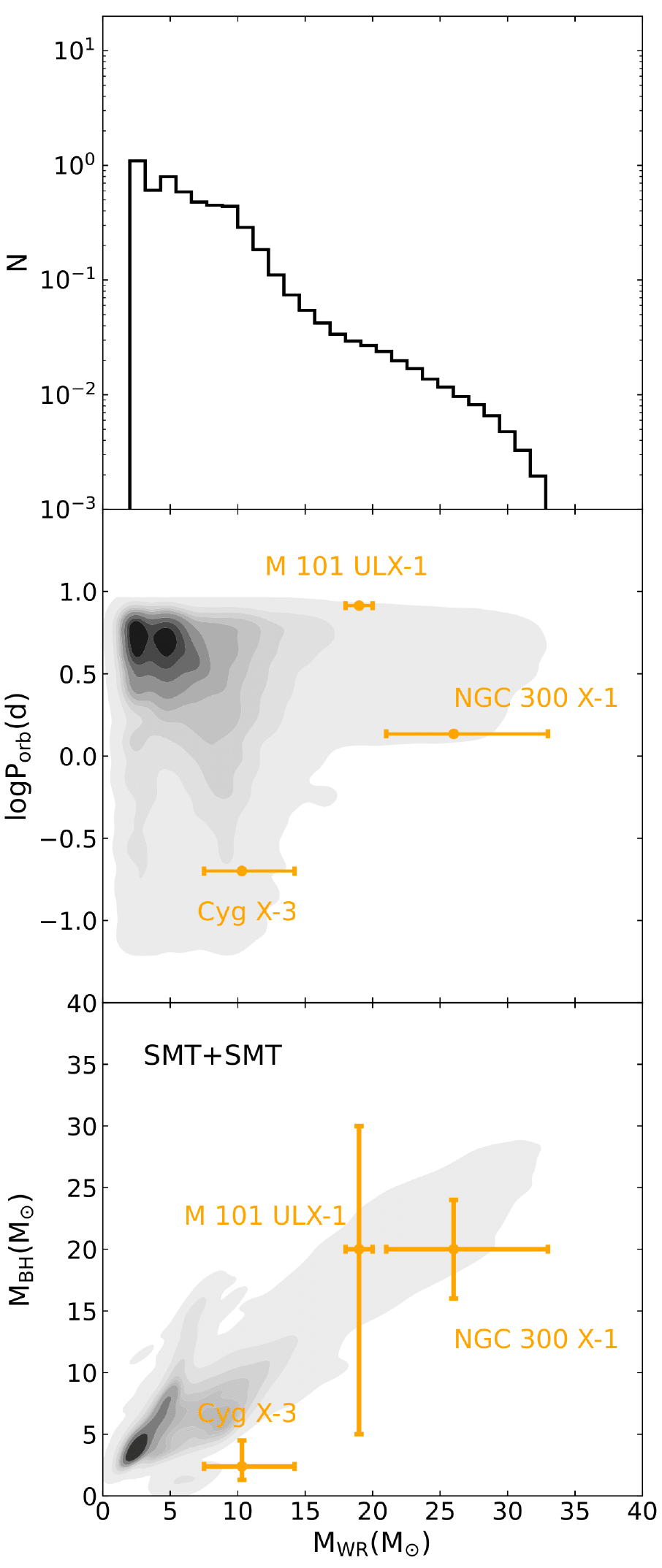}
    \end{minipage}%
    \caption{Predicted number distribution of the BH+WR binaries formed through the SMT+SMT channel, by assuming a constant star formation rate of 0.5 $M_{\odot} \mathrm{yr}^{-1}$ over the past 100 Myr. In the upper panel, the histogram is plotted showing the number distribution as a function of $M_{\mathrm{WR}}$. In the middle and lower panels, gray contours are plotted showing the number distribution in the  $M_{\mathrm{WR}}-P_{\rm orb}$ and $M_{\mathrm{WR}}-M_{\mathrm{BH}}$ planes, respectively. 
    The orange dots mark the positions of three observed BH+WR binaries.}
    \label{fig: MassDALogPSingle100}
\end{figure}

\begin{figure*}[htbp]
  \centering
  \includegraphics[width=0.9\textwidth]{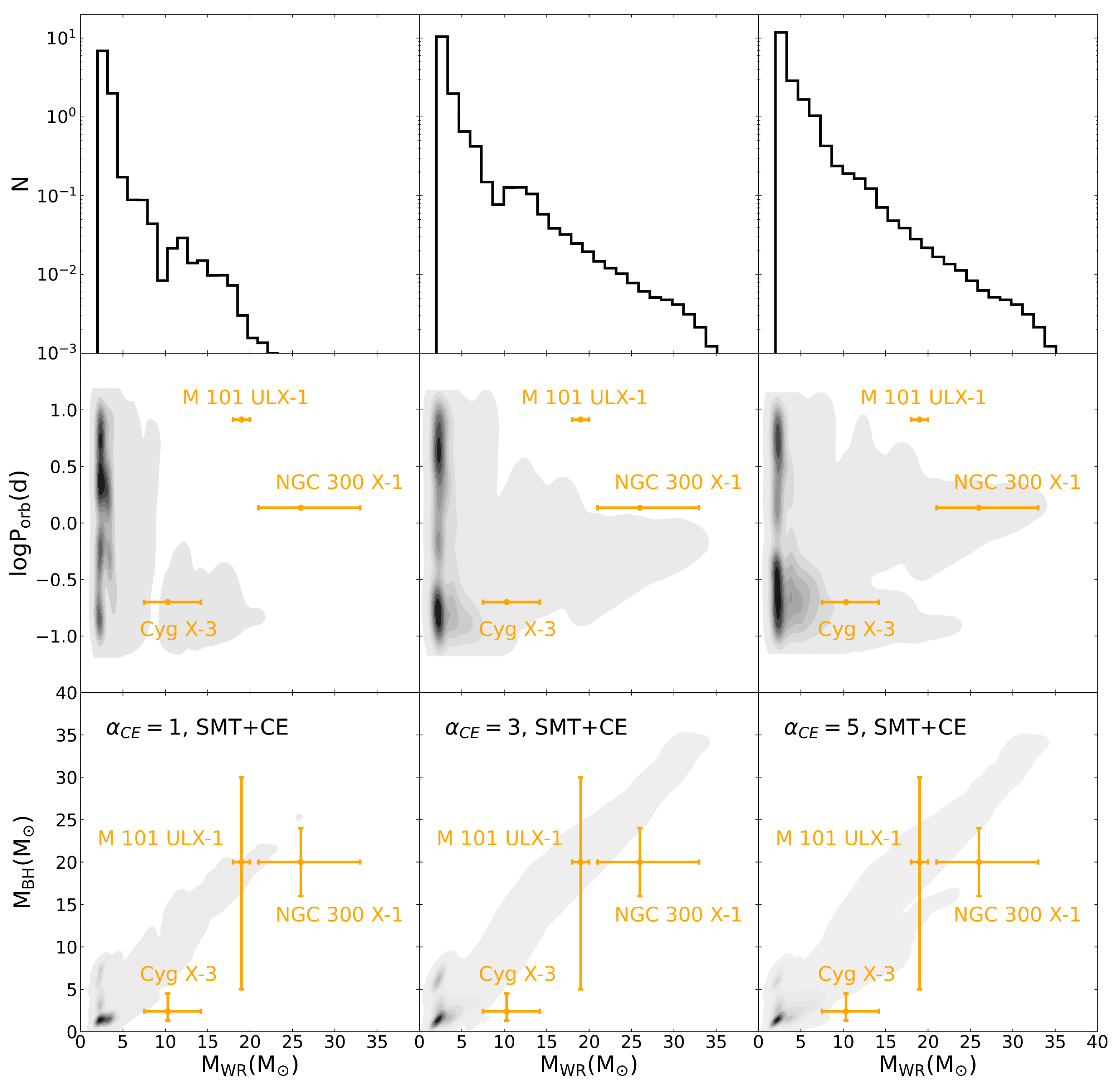}
  \caption{Similar to Figure \ref{fig: MassDALogPSingle100}, but for the SMT+CE channel. The left, middle and right panels correspond to the cases with 
    $\mathrm{\alpha_{\mathrm{CE}}}=1$, 3 and 5, respectively.}
  \label{fig: MassDALogPSMTCE100}
\end{figure*}

\begin{figure*}[htbp]
  \centering
  \includegraphics[width=0.9\textwidth]{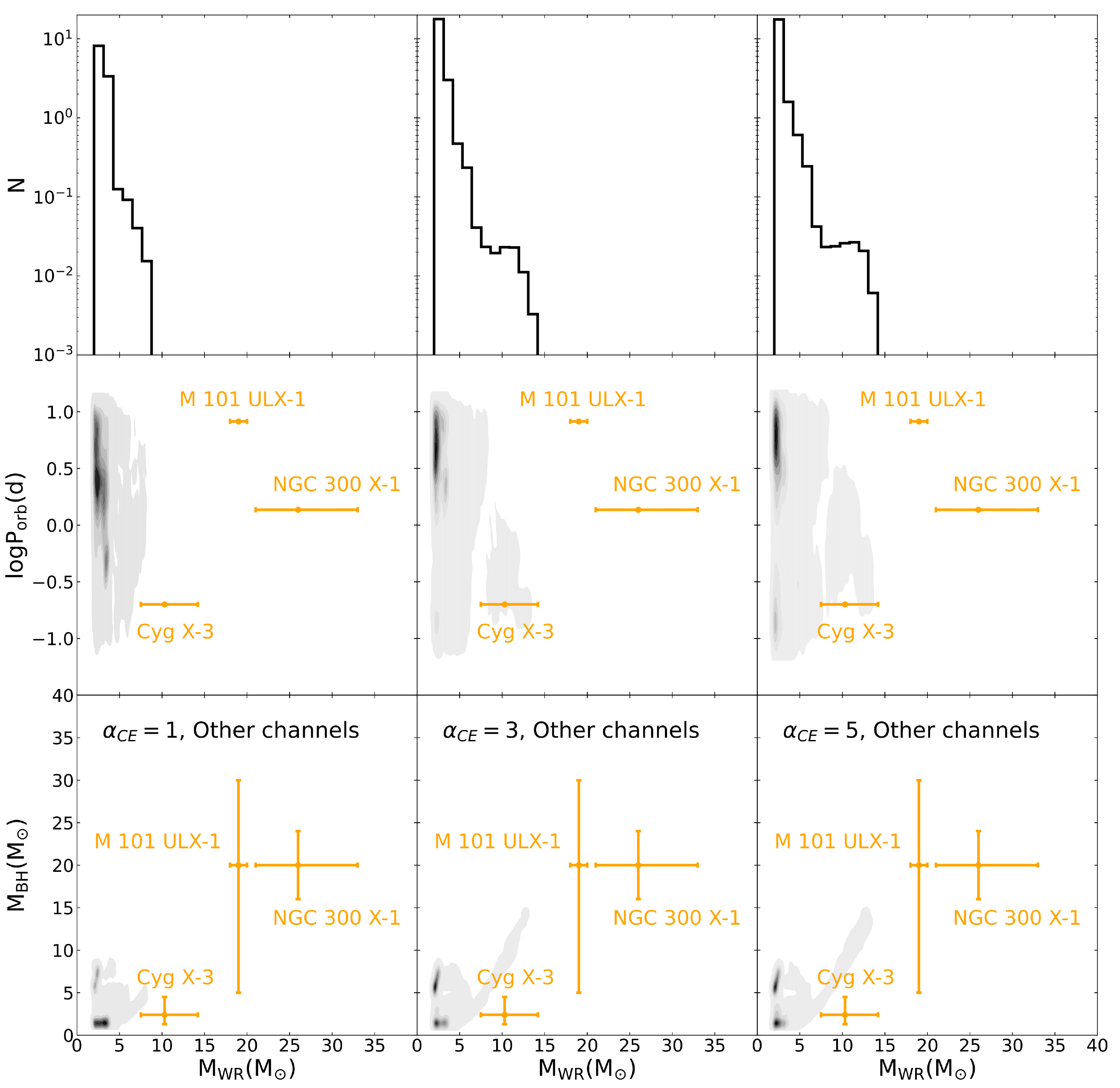}
  \caption{Similar to Figure \ref{fig: MassDALogPSMTCE100}, but for other formation channels excluding SMT+SMT and SMT+CE.}
  \label{fig: MassDALogPOther100}
\end{figure*}

Figures \ref{fig: MassDALogPSingle100}-\ref{fig: MassDALogPOther100} show the number distributions of the BH+WR binaries former via the SMT+SMT channel, the SMT+CE channel and other channels, respectively. Here, we relax the maximal evolutionary age to be 100 Myr. 

        \clearpage
    \bibliographystyle{aasjournal}
    \bibliography{article}

\end{document}